\documentclass{jpsj2}
\usepackage[dvips]{graphics}

\def\runtitle{Numerical study of the 
lattice vacancy effects on ........ }
\def\runauthor{Katsunori \textsc{Wakabayashi}}

\def\mbf(#1){\mbox{\boldmath $#1$}}

\title{%
Numerical study of the 
lattice vacancy effects on the single-channel electron transport
of graphite ribbons}

\author{%
Katsunori \textsc{Wakabayashi}$^{1,2}$
\thanks{Email address: waka@qp.hiroshima-u.ac.jp}
}

\inst{%
$^1$Department of Quantum Matter Science,
Graduate School of Advanced Sciences of Matter (ADSM), 
Hiroshima University, Higashi-Hiroshima 739-8527 \\
$^2$Department of Physical Electronics, Faculty of Engineering, 
Hiroshima University, Higashi-Hiroshima 739-8527 
}

\recdate{\today}

\abst{%
Lattice vacancy effects on electrical conductance 
of nanographite ribbon are investigated by means of the Landauer 
approach using a tight binding model. 
In the low-energy regime ribbons with zigzag
boundary provide a single conducting channel whose origin is connected 
with the presence of edge states. 
It is found that
the chemical potential dependence of conductance strongly depends on the
difference ($\Delta$)  of the number of removed A and B sublattice sites.
The large lattice vacancy with $\Delta\neq 0$ shows
$2\Delta$ zero-conductance dips in the single-channel region,
however, the large lattice vacancy with $\Delta=0$ has no dip
structure in this region.
The connection between this conductance rule and the Longuet-Higgins
conjecture is also discussed.
}

\kword{%
Nanographite, graphite ribbon, lattice vacancy, zero-conductance,
Landauer formula 
}

\begin{document}
\sloppy
\maketitle

%%%%%%%%%%%%%%%%%%%%%%%%%%%%%%%%%%%%%%%%%%%%%%%%%%%%%%%%%%%%%%%%%%%%%%%%%%%%%%
\section{INTRODUCTION}
%%%%%%%%%%%%%%%%%%%%%%%%%%%%%%%%%%%%%%%%%%%%%%%%%%%%%%%%%%%%%%%%%%%%%%%%%%%%%%
Recently, nanometer-sized carbon systems such as carbon
nanotube\cite{review1,review2}, fullerene molecule and
nanographite have attracted much attention 
by the possibilities for the realization
of carbon-based molecular-electronic devices.
In these systems, 
the geometry of sp$^2$ carbon networks
has much influence on the electronic states near the Fermi level.
Not only the closed carbon molecules such as carbon nanotubes and
fullerene molecules, but also systems with open boundaries also display unusual
features connected with their boundaries.
The existence of graphite edges 
strongly affects the $\pi$-electronic states
in nanometer-sized graphite fragments (nanographites).\cite{peculiar} 
There are two basic edge shapes in graphite,
{\it armchair} and {\it zigzag}.
For the model of graphite ribbons, 
one-dimensional graphite lattices of finite
width,\cite{peculiar,nakada,waka}
it was shown that ribbons with zigzag edges (zigzag ribbon) possess 
localized edge states with energies close to
the Fermi level.\cite{peculiar,nakada,waka}
These edge states correspond to the non-bonding molecular
orbitals as can be seen by examining the analytic solution for
semi-infinite graphite with a zigzag edge.
In contrast, edge states are
completely absent for ribbons with armchair edges. 
We have also pointed out that
the edge states play important roles in magnetic properties
in nanometer-sized graphite systems,
because of their relatively large contribution to the density of
states at the Fermi energy.\cite{peculiar,nakada,waka,rpa}
In this paper, we study the impurity scattering effect 
and the lattice vacancy effect in
the electron transport of zigzag nanographite ribbons,
in which the conductance crucially depends on the
geometry of the lattice defects.
We find that the conductance rule:
the large lattice vacancy with $\Delta\neq 0$ shows
$2\Delta$ zero-conductance dips in the single-channel region,
however, the large lattice vacancy with $\Delta=0$ has no dip
structure in that region.

%%%%%%%%%%%%%%%%%%%%%%%%%%%%%%%%%%%%%%%%%%%%%%%%%%%%%%%%%%%%%%%%%%%%%%%%%%%%%%
\section{ ELECTRONIC PROPERTIES OF NANOGRAPHITE RIBBONS }
\begin{figure}
%\begin{minipage}{\textwidth}
\begin{center}
\scalebox{.52}{\includegraphics{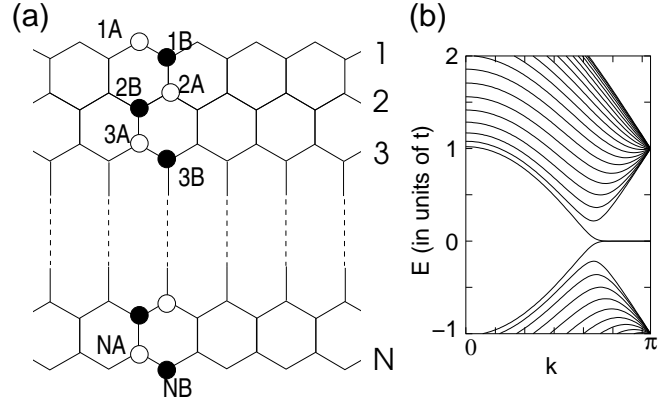}}
%\psbox[width=80mm]{Fig1.eps}
\end{center}
\caption{ (a) Graphite ribbon with zigzag edges (zigzag ribbon). (b)
 Energy band structure of zigzag ribbon (N=20).}
\label{fig.ribbon}
%\end{minipage}
\end{figure}

%%%%%%%%%%%%%%%%%%%%%%%%%%%%%%%%%%%%%%%%%%%%%%%%%%%%%%%%%%%%%%%%%%%%%%%%%%%%%%
Let us introduce the electronic states of
nanographites based on the tight binding model.
The tight-binding Hamiltonian is defined by
\begin{eqnarray}
H = \sum_{i,j} t_{i,j}|i\rangle\langle j| 
  + V\sum_\alpha |\alpha\rangle\langle\alpha |, 
\end{eqnarray}
where $t_{i,j}=-t$ if $i$ and $j$ are nearest neighbors, otherwise 0, 
and $|i\rangle$ is a localized orbital on site $i$.
The second term represents the impurity potential, 
V and $\alpha$ are the strength of the
impurity potential and the location of impurity, respectively.
In Fig.\ref{fig.ribbon}(a), the graphite ribbon with zigzag edges 
(zigzag ribbons ) is shown.
We assume all edge sites are terminated by H-atoms.
The ribbon width $N$ is defined by the number of 
zigzag lines. 
As the graphite lattice is bipartite, the
A(B)-site on the n-$th$ zigzag line is called nA(nB)-site.
Fig. \ref{fig.ribbon}(b) is 
the energy band structure of zigzag ribbon for $N=20$. 
The zigzag ribbons are metallic for all $N$.
One of the remarkable features is the appearance of
partly flat bands at the Fermi level($E=0$), 
where the electrons
are strongly localized near zigzag edges \cite{peculiar}.

The analytic solution of the partly flat band
for a semi-infinite graphite sheet with a zigzag edge 
( 1A and 1B site shall be the edge sites and N is infinite
in Fig.\ref{fig.ribbon}(a)) can be expressed as 
$\phi_{nA} = D_k^{n-1}$ and  $\phi_{nB} = 0$,
where $\phi_{nA}$  ($\phi_{nB}$) means the amplitude of the
edge states on nA (nB) site and $D_k = -2\cos(k/2)$.
It is worth noting that the edge state has 
a non-zero amplitude
only on one sublattice, i.e.  non-bonding character.
Because of the convergence condition of the edge states, 
the wave number $k$ must be $2\pi/3 \le k\le \pi$,
where $|D_k|\le 1$.
In this $k$-region, the edge states make a flat band  
at E=0 (Fermi energy).
It should be noted that at $k=\pi$ the edge states are perfectly
localized at the 1A sites, but  at $k=2\pi/3$ the edge states are
completely delocalized.
However, when we consider the zigzag ribbons,
two edge states which come from both sides will overlap 
each other and cause the bonding and anti-bonding splitting.
The magnitude of the overlap becomes larger when the wave number 
approaches $2\pi/3$, because the penetration length of the
edge states gets larger there.
Therefore, the partly flat bands have a slight dispersion, which
leads to one channel for the electron transport.
The energy region of the single-channel, $\Delta_z$, ( defined by the 
energy gap to the next channel) is approximately,
$\Delta_z=-4t\cos\left[(N-1)\pi/(2N+1)\right]$.

In this manuscript, we evaluate
the electrical conductance by the Landauer formula\cite{mclbf},
\begin{eqnarray}
G(E) = \frac{e^2}{\pi\hbar}T(E),
\end{eqnarray}
where $T$ is the transmission probability through the
impurity potential. 
We use the recursive Green function method for the calculation of
the transmission probability\cite{MacKinnon}, which provides
high numerical accuracy and efficiency.
We also use this method to calculate
the electron waves and electric current 
around impurity\cite{prl,prb2}.

%%%%%%%%%%%%%%%%%%%%%%%%%%%%%%%%%%%%%%%%%%%%%%%%%%%%%%%%%%%%%%%%%%%%%%%%%%%%%%
\section{ SINGLE IMPURITY SCATTERING }
%%%%%%%%%%%%%%%%%%%%%%%%%%%%%%%%%%%%%%%%%%%%%%%%%%%%%%%%%%%%%%%%%%%%%%%%%%%%%%
In this section, we study the single impurity scattering effect in the
nanographite ribbons. 
In the low-energy single-channel region,
even the existence of a single-impurity causes the 
the non-trivial behavior to the
position and strength dependence of the impurity due to the
edge localized nature of the edge states.

Before we consider the electron transmission through 
an impurity or lattice vacancy in zigzag ribbon,
let us discuss the single impurity problem in a graphite sheet.
The problem described here is the electron scattering by the 
$\delta-$function type impurity potential with the strength $V$,
at an A-site ( $\mbf(r)=0$ ). The Schr\"{o}dinger equation is written as
\begin{eqnarray}
\left(
\begin{array}{cc}
V\delta\left({\bf r}\right) & \epsilon_k^\ast \\
 \epsilon_k  &  0
\end{array}
\right)
\left(
\begin{array}{c}
\phi_{\rm A}({\bf r})\\ \phi_{\rm B}({\bf r})
\end{array}
\right) = E
\left(
\begin{array}{c}
\phi_{\rm A}({\bf r})\\ \phi_{\rm B}({\bf r})
\end{array}
\right),
\end{eqnarray}
where $\epsilon_k=-t\left(
\exp(-ik_ya) 
+ \exp(i( \sqrt{3}k_x + k_y)a/2
+ \exp(-i(\sqrt{3}k_x - k_y)a/2
\right)$ describes the electron hopping 
from the A site to the B site in the unit cell.
Here we take the translational invariant direction of the zigzag axis 
as the x-axis, the y-axis is perpendicular to the x-axis, and
$ a $ is the lattice constant.
Following the standard approach by Koster and Slater\cite{koster},
we obtain the self-consistent equation determining the
impurity state energy, $E$, as follows,
\begin{eqnarray}
\frac{1}{V}=\displaystyle{
\frac{E}{(2\pi)^2} 
\int_{\rm\bf 1st BZ} d\mbf(k)
\frac{1}{E^2- |\epsilon_k|^2 }},
\end{eqnarray}
where the integral is taken over the 1st Brillouin zone of the
graphite sheet. 
This integral is in general complex number, which means
the bound state has a resonant nature, i.e. virtual bound state
with finite lifetime.
However, in the limit of infinity $V$, i.e. a single vacancy in the
graphite sheet, the virtual bound state becomes well-defined
state bound to the impurity, and the energy level of
the impurity state comes to $E=0$.
We should note that this resonant behavior is quite similar to
the single impurity problem in a $d$-wave superconductor\cite{balatsky},
because of the finite density of states 
$D(E)\propto E$ in both systems. 
The wavefunction then can be written in $\mbf(k)$-space as
\begin{eqnarray}
\phi_{Ak} = & \displaystyle{\frac{V}{L^2} G^0_A \frac{E}{E^2- |\epsilon_k|^2 }} \\
\phi_{Bk} = & \displaystyle{\frac{V}{L^2} G^0_A \frac{\epsilon_k}{E^2- |\epsilon_k|^2 }},
\end{eqnarray}
where $G^0_A=\sum_k \phi_{Ak}$.
From the normalization of these wavefunctions, we can find that,
in the limit of infinity $V$, $\phi_A(\mbf(r))=0$ and
$\phi_B(\mbf(r))\neq 0$, i.e. non-bonding character.
Thus if we make a single vacancy at an A(B) site
in a graphite sheet, a vacancy bound state appears at $E=0$
and its wavefunction has a non-bonding character,
where the amplitudes of the A(B)-sites are zero.

\begin{figure}
%\begin{minipage}{\textwidth}
%\begin{center}
\scalebox{.52}{\includegraphics{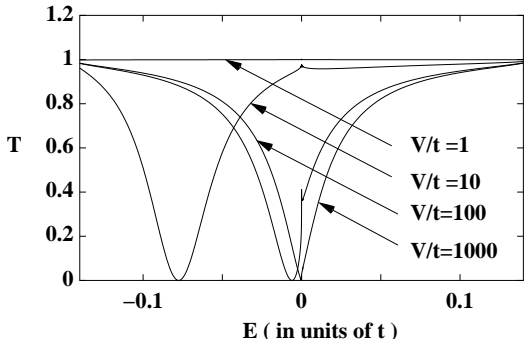}}
%\psbox[width=80mm]{Fig2.eps}
%\end{center}
\caption{Fermi energy dependence of the transmission
probability for the various strength of impurity potential
($V/t=1, 10^1, 10^2, 10^3$)
when the impurity is located at 15B site of
the zigzag ribbon with N=30.}
\label{fig.single.imp}
%\end{minipage}
\end{figure}

In Fig.\ref{fig.single.imp}, we show the Fermi energy dependence of the
transmission  
probability in the single-channel energy region
for the graphite ribbon with $N=30$, where
the impurity is located at center of the ribbon (15B site).
The strength of the impurity potential is changed by
$V/t=1, 10^1, 10^2, 10^3$. 
The conductance for negative $V/t$ can be obtained by the 
transformation of $E\rightarrow -E$ in Fig.\ref{fig.single.imp}.
We find pronounced conductance dip structures for sufficiently strong
impurity potential. 
At the energies of conductance dips, the electron waves
are localized around the impurity site. Thus the conductance
dips is caused by virtual bound states at the impurity.
We see also that
the phase jumps of the transmission coefficient by $\pi$ at 
the energy of dips. The $\pi$-phase jump guarantees the perfect
reflection at the conductance dip\cite{prl}.
The weak impurity potential (less than $t$)
does not have much influence on the conductance in the single-channel
region, although the graphite ribbon is one-dimensional system.
The width of zero-conductance resonance means the
inverse of the traversal time that the electrons pass through the 
impurity region.
In many cases, the traversal time increases with 
apart from $E=0$, 
because the group velocity of edge states gets larger
with leaving from $E=0$. Thus the conductance drastically changes 
by the strength of impurity potential.

\begin{figure}
%\begin{minipage}{\textwidth}
%\begin{center}
\scalebox{.52}{\includegraphics{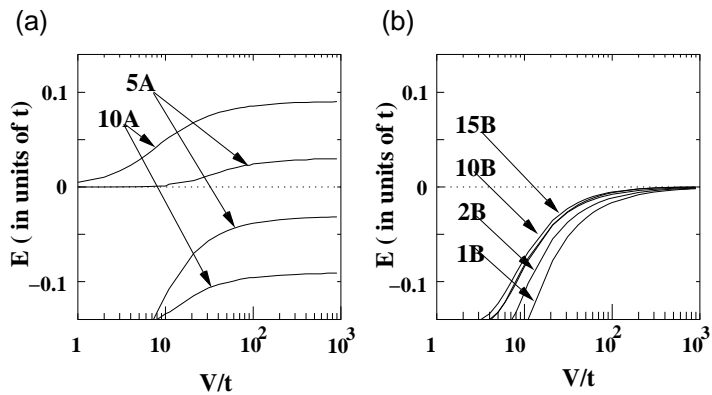}}
%\psbox[width=100mm]{Fig3.eps}
%\end{center}
\caption{The dependence of the position of zero-conductance dips
on the strength of impurity potential (a) for 5A and 10A site, 
(b) for 1B, 2B, 10B and 15B.}
\label{fig.single.vdep}
%\end{minipage}
\end{figure}

In Fig.\ref{fig.single.vdep}, we show the dependence of the position of
zero-conductance dip
on the strength of impurity potential. In the limit of
$V/t\rightarrow\pm\infty$, the impurity level approaches  $E=0$, 
where a single nA(B)-site impurity produces a non-bonding state
so that the amplitude of nA(B)-sites is zero.
On the other hand, 
the edge states has the character that electrons strongly
localized on nA(nB)-sites for n$<N/2$ (n$>N/2$).
Therefore the impurity state caused by the impurity potential
located at nA-site(n$<N/2$) strongly couples
with the edge states, resulting in the splitting of the level of
zero-conductance dips(Fig.\ref{fig.single.vdep}(a)). However,
the impurity state caused by the impurity potential located at
nB-site(n$<N/2$) cannot couple with the 
edge states because the impurity states and edge states live
on the same sublattice sites, so that zero-conductance dip
appears only near $E=0$(Fig.\ref{fig.single.vdep}(b)) in the strong
$V/t$ limit. 
Note that nB-site impurity does not produce 
zero-conductance dips in weak $V/t$ region, and
nA-site impurity produces a single zero-conductance dip
close to $E=0$ in weak $V/t$ region.

\begin{figure}
%\begin{minipage}{\textwidth}
%\begin{center}
\scalebox{.52}{\includegraphics{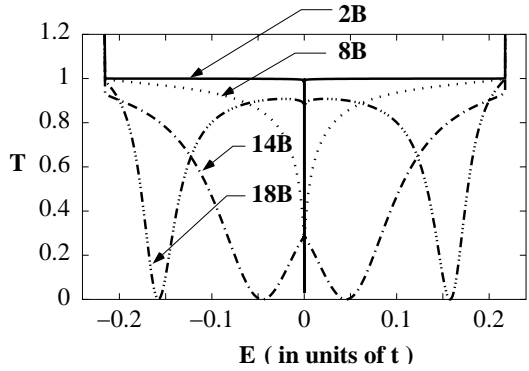}}
%\psbox[width=60mm]{Fig4.eps}
%\end{center}
\caption{Fermi energy dependence of the transmission
probability of zigzag ribbon (N=20) with a single-site vacancy.
The position of the vacancy site is denoted in the figure.}
\label{fig.single.vacan}
%\end{minipage}
\end{figure}

The Fermi energy dependence of the transmission through
a single vacancy($V/t\rightarrow\infty$) in zigzag ribbon is shown in 
Fig.\ref{fig.single.vacan}.
The position of the vacancy site is denoted in the figure.
In this limit, two zero-conductance dips appear with mirror symmetry to
$E=0$. Since a single vacancy on nA(B)-site
produces a zero-energy non-bonding state 
with zero-amplitude on the nA(B)-sites,
the non-bonding states at the vacancy and one of the two edges couple 
forming a bonding and antibonding configuration with energy level
below and above zero causing the zero-conductance dips at the
corresponding energies.  
The vacancy state due to nB-site vacancy for n$<N/2$ are so far removed
from the corresponding edge that their overlap is very small
and the corresponding zero-conductance dips occur at energy extremely
close to $ E=0 $. 
Thus, in the low-energy single-channel region,
the electric conductance shows the non-trivial behavior, 
even for the single impurity case, due to the 
edge localized non-bonding nature of the edge states and
non-bonding character of the impurity states.

Here we briefly discuss the origin of the zero-conductance
dips observed. 
Each zero-conductance resonance can be associated with
a quasi-bound state around the vacancy or the impurity, yielding
the formation of standing waves.
Around the energies of the each zero-conductance resonances,
the electric currents show the Kekul\'{e}-like current
vortex pattern, which can be observed in the
nanographite ribbon junctions.\cite{prl,prb2}
We have also checked each resonance feature can be characterized by
a zero-pole pair in the complex energy plane.
This suggest that the zero-conductance dip corresponds to a
Fano resonance which is known
to occur when two scattering channels are available, one corresponding
to a continuum of states and the other to a discrete quasi-bound
state.\cite{shao,porod}

%%%%%%%%%%%%%%%%%%%%%%%%%%%%%%%%%%%%%%%%%%%%%%%%%%%%%%%%%%%%%%%%%%%%%%%%%%%%%%
\section{ LARGE SIZE LATTICE VACANCY }
%%%%%%%%%%%%%%%%%%%%%%%%%%%%%%%%%%%%%%%%%%%%%%%%%%%%%%%%%%%%%%%%%%%%%%%%%%%%%%

\begin{figure}
\scalebox{.52}{\includegraphics{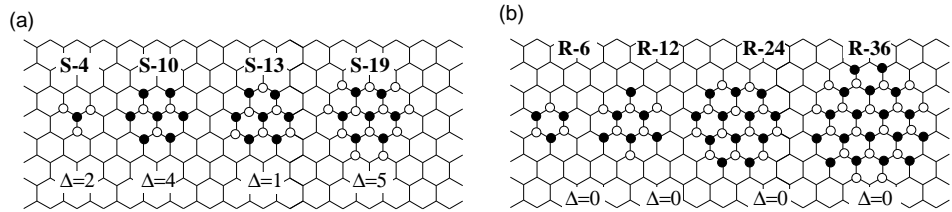}}
\caption{Examples of (a) site-centered and (b) ring-centered vacancies, and
$\Delta$ denotes the difference of the number of removed A and B
sublattice sites.}
\label{fig.vacan.struct}
\end{figure}

Next we study the effect of large size lattice vacancies.
Examples of large size lattice vacancy are shown in
Fig.\ref{fig.vacan.struct}, where 
the removed sites are denoted by black and white circles.
There are two typical lattice vacancies, (i) site-centered vacancies
(Fig.\ref{fig.vacan.struct}(a)) and (ii) ring-centered vacancies
(Fig.\ref{fig.vacan.struct}(b)). 
They have three- and six-fold symmetry, respectively.
The notation S-m (or R-m) denotes the
site(ring)-centered lattice vacancy 
with m removed sites.
The difference $\Delta=|N_A - N_B|$ is
always zero for ring-centered vacancy, where
$N_A$ ($N_B$) is the number of removed A (B) sublattice 
sites.

\begin{figure}[hbt]
%\begin{minipage}{\textwidth}
%\begin{center}
\scalebox{.52}{\includegraphics{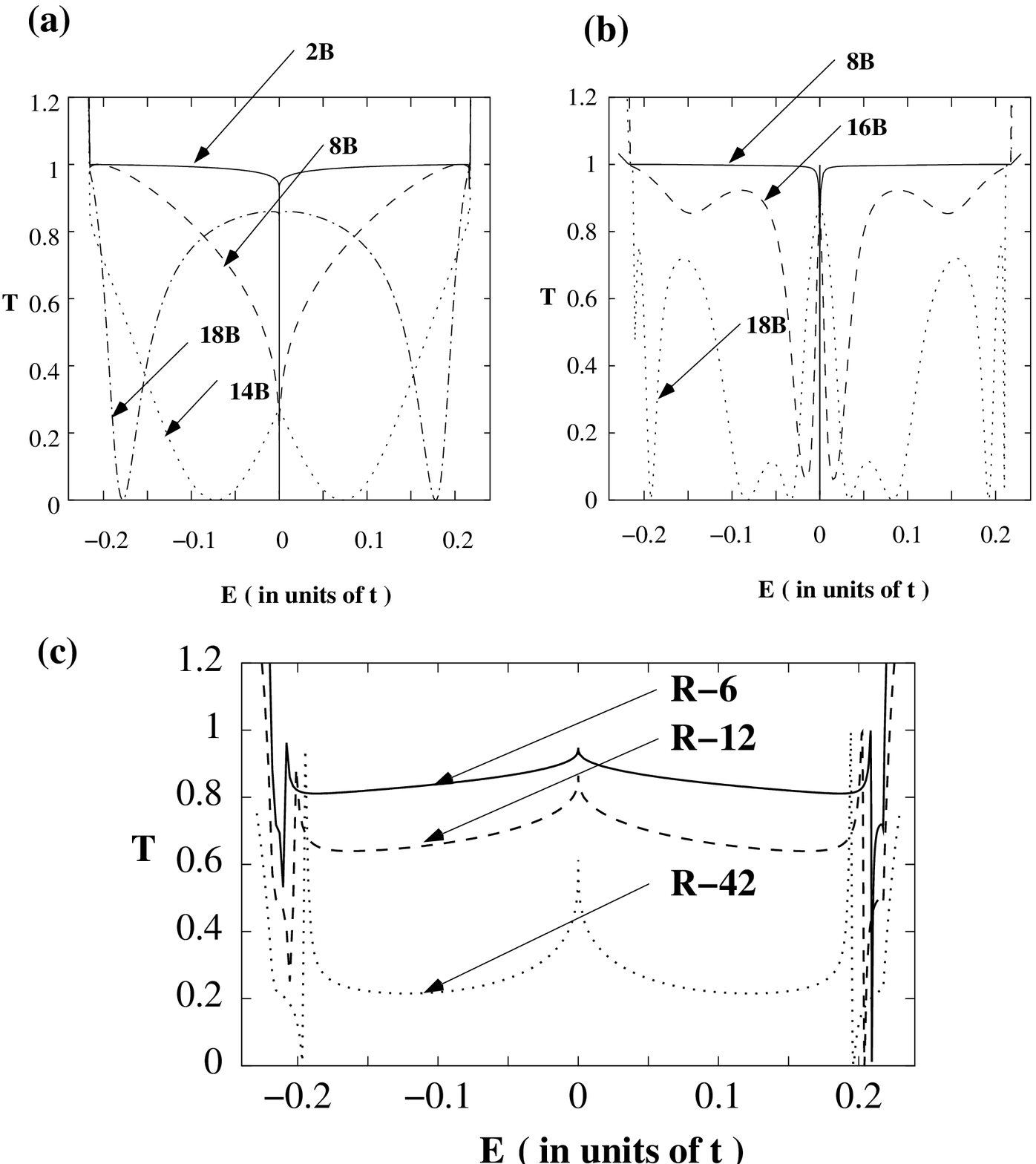}}
%\psbox[width=90mm]{Fig6.eps}
%\end{center}
\caption{Fermi energy dependence of the transmission probability
for the site-centered vacancies of (a) S-13 and (b) S-10, where 
the indices denote the center position of the vacancy.
(c) The transmission probability for ring-centered vacancies 
of R-6, R-12 and R-42, which are located at the center of the
ribbon.}
\label{fig.vacan.conduc}
%\end{minipage}
\end{figure}

The conductance of zigzag ribbon (N=20) with site-centered vacancies
is shown in Fig.\ref{fig.vacan.conduc} for the case of (a) S-13 and (b) S-10.
The indices in the figure mean the center position of
the vacancy. 
The conductance is very sensitive to the position
of vacancy, similar to the single-site vacancy case.
In analogy to the Longuette-Higgins(LH) conjecture\cite{LH},
the vacancy with $\Delta\ne 0$($N_A<N_B$)
produces $\Delta$-degenerate
non-bonding vacancy states where the amplitude of 
nB-sites is zero.
Therefore, the S-13 vacancy ($\Delta(=N_B-N_A)=1$) 
produces a non-bonding vacancy state, where the nB sites are
node, and couples with the edge state which comes from the upper edge in
Fig. \ref{fig.ribbon}(a). 
The case of the S-10 vacancy which is $\Delta=4$ produces, however,
many more zero-conductance dips, with increasing the strength of
coupling between the edge states and vacancy states.
Therefore it is considered that, if the large size
single vacancy with $\Delta\neq 0$ such as the Fig.5 is introdunced
in a graphite sheet, the $\Delta$-degenerated vacancy bound states 
are produced at $E=0$.

In Fig.\ref{fig.vacan.conduc} (c), the Fermi energy dependence of
the conductance of zigzag ribbon (N=20) with a
ring-centered vacancy is shown.
The conductance gradually decreases with increasing 
the size of the vacancy, however
no dip structure appears in the single-channel region.
The conductance dips appear only around the energy where
the channel number changes. 
Since these dips are caused by the multi-channel scattering effects,
we do not pay attention to these dip structures in this paper.
Since in analogy to the LH conjecture, 
the vacancy with $\Delta=0$ does not create non-bonding vacancy states,
the zero-conductance resonances do not appear in the
single-channel energy region.
Interestingly, 
the conductance is not sensitive to the position of
ring-centered vacancy. 
The resonant tunneling-like behavior around $E=0$ is mimic,
and is caused by the variation of the group velocity,
because the group velocity rapidly
decreases with approaching to $E=0$ due to the partly flat band.
It should be noted that the transmission probability is exactly zero and
singular at $E=0$, because the group velocity is zero at $E=0$.

Thus our numerical results show that the large
size vacancy with $\Delta\neq 0$ produces $2\Delta$ 
zero-conductance dips in the single-channel energy region,
however, the large size vacancy with $\Delta = 0$
has no zero-conductance dip in the single-channel region.
This rule is not strict, so that there are a few exceptions 
which do not satisfy the rule because of the complicated
multiple-scattering effects between edge states and
vacancy states.
However, most of the vacancies numerically tested satisfy this
conductance rule.
Furthermore, it should be noted that this conductance rule
implies the following rule: the large size vacancy with  $\Delta\neq 0$
in a graphite sheet produces the $\Delta$-degenerated
zero-energy non-bonding bound states and
the vacancy with $\Delta=0$ in a graphite sheet
do not produce the bound state.
Thus this rule has the negative-positive relation with the
LH-rule discussed in a finite molecule.
A rigorous proof remains a problem for future study.

%%%%%%%%%%%%%%%%%%%%%%%%%%%%%%%%%%%%%%%%%%%%%%%%%%%%%%%%%%%%%%%%%%%%%%%%%%%%%%
\section{\bf CONCLUSION}
%%%%%%%%%%%%%%%%%%%%%%%%%%%%%%%%%%%%%%%%%%%%%%%%%%%%%%%%%%%%%%%%%%%%%%%%%%%%%%

In conclusion, we have studied the effects of a non-magnetic impurity on
electrical conductance of graphite ribbons with zigzag edges.
The conductance strongly depends on the position and strength of
the impurity. For large size lattice vacancy,
the behavior of the conductance drastically changes with the
difference ($\Delta$) of removed A and B sublattice sites.
For the $\Delta\ne 0$ case, the Fermi energy dependence of
conductance shows dip structures.
For the $\Delta= 0$ case, conductance dips appear only
around the energies where the number of channel changes,
but no conductance dips in the single channel region.
Our results support that the lattice vacancy with $\Delta\neq 0$
produces the $\Delta$-non-bonding vacancy bound states, however
the vacancy with $\Delta=0$ do not produce bound state.
Our work shows that the impurity effects in 
nano-graphite ribbons are quite different from the case of
usual quantum wires\cite{bagwell} and carbon nanotubes\cite{igami}.
In this paper, we have pointed out that the electric conductance
through the nanographite ribbons with a single vacancy 
produces the pronounced zero-conductance resonances 
and shows crucial dependence on the vacancy geometry and the strength of
the impurity potential. Therefore the nanographite ribbons which posses
many vacancy or impurity potential show the non-trivial behavior
in the electrical conductance. We demonstrate this problem elsewhere.

%%%%%%%%%%%%%%%%%%%%%%%%%%%%%%%%%%%%%%%%%%%%%%%%%%%%%%%%%%%%%%%%%%%%%%%%%%%%%%
\section{ Acknowledgments}
%%%%%%%%%%%%%%%%%%%%%%%%%%%%%%%%%%%%%%%%%%%%%%%%%%%%%%%%%%%%%%%%%%%%%%%%%%%%%%
I would like to thank M. Sigrist, M. Igami and T. Kawabe
for their many helpful discussions.
I am grateful for support by Grant-in-Aid for Scientific
Research from Ministry of Education, Science and Culture, Japan,
and the financial support from the Foundation Advanced Technology 
institute and from the Kinki-chihou Hatsumei Center.
Numerical calculations were performed in part in 
Institute for Molecular Science.

%%%%%%%%%%%%%%%%%%%%%%%%%%%%%%%%%%%%%%%%%%%%%%%%%%%%%%%%%%%%%%%%%%%%%%%%%%
%%%%%%%%%%%%%%%%%%%%%%%%%%%%%%%%%%%%%%%%%%%%%%%%%%%%%%%%%%%%%%%%%%%%%%%%%%

%%%%%%%%%%%%%%%%%%%%%%%%%%%%%%%%%%%%%%%%%%%%%%%%%%%%%%%%%%%%%%%%%%%%%%%%%%
%%%%%%%%%%%%%%%%%%%%%%%%%%%%%%%%%%%%%%%%%%%%%%%%%%%%%%%%%%%%%%%%%%%%%%%%%%

%%%%%%%%%%%%%%%%%%%%%%%%%%%%%%%%%%%%%%%%%%%%%%%%%%%%%%%%%%%%%%%%%%%%%%%%%%
%%%%%%%%%%%%%%%%%%%%%%%%%%%%%%%%%%%%%%%%%%%%%%%%%%%%%%%%%%%%%%%%%%%%%%%%%%

%%%%%%%%%%%%%%%%%%%%%%%%%%%%%%%%%%%%%%%%%%%%%%%%%%%%%%%%%%%%%%%%%%%%%%%%%%
%%%%%%%%%%%%%%%%%%%%%%%%%%%%%%%%%%%%%%%%%%%%%%%%%%%%%%%%%%%%%%%%%%%%%%%%%%

%%%%%%%%%%%%%%%%%%%%%%%%%%%%%%%%%%%%%%%%%%%%%%%%%%%%%%%%%%%%%%%%%%%%%%%%%%
%\begin{minipage}{\textwidth}
%\end{minipage}
%%%%%%%%%%%%%%%%%%%%%%%%%%%%%%%%%%%%%%%%%%%%%%%%%%%%%%%%%%%%%%%%%%%%%%%%%%

%%%%%%%%%%%%%%%%%%%%%%%%%%%%%%%%%%%%%%%%%%%%%%%%%%%%%%%%%%%%%%%%%%%%%%%%%%
%%%%%%%%%%%%%%%%%%%%%%%%%%%%%%%%%%%%%%%%%%%%%%%%%%%%%%%%%%%%%%%%%%%%%%%%%%


\begin{thebibliography}{99}
\bibitem{note}
Note that graphite ribbons can be a simple model of
defective carbon nanotubes of carpet-roll or 
papier-m\^{a}ch\'{e}
forms\cite{zhou},
activated carbon fibers\cite{nakayama} 
and nanographites\cite{oeandersson}.
There are also some trials to fabricate nanographite ribbons on
metal substrates\cite{oshima}.
\bibitem{gap}
The zigzag (bearded) ribbons at $k=\frac{2\pi}{3a}$ can be mapped to 
a finite one-dimensional chain with $2N$ ($2N+1$) sites,
which can be solved analytically (see the appendix of \cite{waka}).
From this analytic solution, the form of $\Delta_z$ ($\Delta_b$) is easily
derived.
\end{thebibliography}
\end{document}